\documentclass[
 reprint,
superscriptaddress,
 amsmath,amssymb,
 aps,
pre,
]{revtex4-2}

\usepackage{graphicx}
\usepackage{dcolumn}
\usepackage{bm}

\usepackage{braket}
\usepackage{soul}
\usepackage{xcolor}

\begin{document}
\raggedbottom

\title{Floquet spectra and parameter sensitivity of \\
entanglement in the quantum kicked top}

\author{Dyari Jarvis}
\affiliation{Trinity College, University of Cambridge, Cambridge, United Kingdom}

\author{Peter A. Rose}
\affiliation{Department of Physics, University of Ottawa, Ottawa, Ontario K1N 6N5, Canada}

\author{Arjendu K. Pattanayak}
\affiliation{Department of Physics and Astronomy, Carleton College, Northfield, Minnesota 55057, USA}

\date{\today}

\begin{abstract}

We investigate the relationship between the Floquet spectrum and the sensitivity of the entanglement landscape to parameter and initial condition change in the quantum kicked top. 
We focus on the case of four qubits and on locating degeneracies in its Floquet eigenphases using the resultant of its characteristic polynomial $p$ and its derivative $\partial_{\lambda}p$ and quantitatively map the sensitivity of the entanglement landscape to changes in kicking strength over the sphere of spin coherent initial states.
We find that prominent sensitivity maxima coincide with either Floquet eigenphase degeneracies or kick strengths exhibiting short-period quantum revivals, with some points satisfying both conditions. These results establish a connection between the spectral structure of the Floquet operator and entanglement landscape sensitivity, demonstrating the utility of the resultant in studying the dynamics of small-$j$ quantum kicked tops. Since this structure extends to all values of $j$, the central lessons apply completely generally.
\end{abstract}

\maketitle


\section{Introduction}

Nonlinear many-body quantum systems exhibit a wide range of transport phenomena, including complex entanglement generation, operator scrambling, and thermalization dynamics. These behaviors are often difficult to characterize: even for initially pure states, both the rate and degree of entanglement generation display a rich dependence on initial conditions and system parameters that remains to be fully understood or exploited. Recent approaches have sought to quantify these processes using measures of information transport such as the entanglement entropy, out-of-time-order correlators (OTOCs), and indicators of eigenstate thermalization (ETH).

Much of this work has focused on distinguishing quantum dynamical regimes through comparison with their classical analogs by studying how the presence or absence of classical chaos correlates with quantum behavior. A paradigmatic model for exploring this correspondence is the quantum kicked top (QKT) \cite{haake_1987}, a collective spin system of total angular momentum $j$ (representing $N=2j$ qubits) subjected to a constant and a periodically applied field with the time-dependent form of a many-to-many Ising model as
\begin{equation}
H = \frac{\pi}{ 2\tau} \hat{J}_y + \frac{\kappa}{2j} \hat{J}_z^2 \sum_{n = -\infty}^{\infty}{\delta (t-n\tau)},
\end{equation}
where we have taken $\hbar =1$ and a specific choice of rotation angle $\pi/2$ as is standard. The first term generates a rotation about the y axis, while the second generates a nonlinear twisting kick through  $\hat{J_z^2}$. The kick strength $\kappa$ parametrizes the nonlinearity, while $\tau$ is the time between successive kicks. 
Owing to the periodic driving, the dynamics may be considered stroboscopically at times $T_n = n\tau$. The system is most conveniently studied via the (unitary) Floquet operator 
\begin{equation}
\hat{U}  = \exp \left( -i \frac{\kappa}{2j}\hat{J}_z^2 \right)\exp \left( -i \frac{\pi}{2}\hat{J}_y \right) = U_{\text{kick}}U_{\text{drift}}.
\end{equation}
The QKT has served as a testbed for studying the relationship between entanglement and classical trajectory chaos \cite{Lakshminarayan_2001, Chaudhury2009, Ghose_2008, Lombardi_ent_2011, Kumari_2018, Lerose_2020}, as well as for exploring distinctly quantum effects that arise in the few-body regime \cite{Ruebeck_2017, Madhok_Dogra_Lakshminarayan_2018, Dogra_2019, Bhosale2018}. In such a periodically driven systems, uniquely quantum signatures are reflected in the properties of the Floquet eigenstates $|\lambda_i\rangle$ and their corresponding eigenphases $\lambda_i$ discussed more in detail below.

Ruebeck \textit{et al.} \cite{Ruebeck_2017} showed that for small $j$, the Floquet eigenphases—initially evenly spaced around the unit circle at zero nonlinearity ($\kappa=0$)—evolve with $\kappa$ and return periodically to their original configuration as $\kappa \to \kappa + 4\pi j$. Similarly, Bhosale \textit{et al.} \cite{Bhosale2018} demonstrated that correlation measures such as entanglement entropy and quantum mutual information exhibit periodicity in $\kappa$, typically with period $2\pi j$. These periodic behaviors are purely quantum in origin and have no classical analogue; they signal a breakdown of the classical correspondence principle and identify domains of non-classical coherence and tunneling. As $\kappa$ increases, the Floquet eigenphases move around the unit circle at different rates leading to avoided crossings, bunching, or near-uniform spacing for particular values of the parameter $\kappa$. Such global spectral structures correspond to qualitative changes in the quantum dynamics. For example, when the eigenphases become evenly spaced, the Floquet operator itself becomes periodic, producing resonant quantum dynamics analogous to quantum resonances in kicked-rotor systems.

In this paper, we show that the Floquet spectral structure governs the entanglement landscape sensitivity in the quantum kicked top as well. In particular, we focus on the four qubit kicked top and study its Floquet eigenphase bunching, and especially the existence of degeneracies and avoided crossings via the resultant of its characteristic polynomial $p$ and its derivative $\partial_{\lambda}p$. We find that the entanglement landscape is highly sensitive to
changes in kick strength, $\kappa$, when $\kappa$ is close to special values.  All of these special values fall into two categories: they are either locations where the eigenphases are equally spaced, and thus the Floquet operator is periodic, or where there are
zero-crossings in the eigenspectrum. Most but not all (the exception being one of the zero-crossings) of these exceptional $\kappa$ values occur at quantum resonances of recent interest in the QKT. These quantum resonances are a  generalization of quantum resonances in the QKR\cite{Santhanam_2022}. In these resonant conditions, the unusual quantum dynamics arises from a coherent repeated application of the Floquet operator, and exist for all $j$ (although the specifics and the details are complicated and differ for $N$ even or odd). Dynamics at specific `resonant' parameters $(r/s)4\pi j$ have been shown to display either classical-like (such as at $\kappa=0$ or small $\kappa$ \cite{Zou_Wang_pseudo_2022}) behavior, or else uniquely quantum behavior such as exact periodicity. The $\kappa$ periodicity of the Floquet operator as $\kappa \to \kappa + 4\pi j$ can be understood as a primary resonance. Recent work\cite{anand_2024, anand_2026} examining recurrences in Floquet dynamics show that these occur at certain quantum resonances (where the spectra correspond to the roots of unity, as we discuss in more detail below); this combination of periodicity and high entanglement has been shown to lead to valuable protocols\cite{zou_2025_enhancing} in quantum metrology. 

Our results thus establish a clear way of locating interesting spectral features of the Floquet operator that are correlated with unusual or parametrically-sensitive entanglement and other dynamical quantum properties that do not depend in any way on invoking the classical or semi-classical limit, as has been typically evoked. These results and behaviors persist in the semiclassical limit, showing deeply quantum dynamics far into this regime, albeit at very finely-tuned parameters, as we discuss below. 
We present all this below starting by describing some details of the model as well as the metrics used after which we present our results and analysis before concluding with some general discussion.

\section{Quantum Kicked Top and entanglement landscape }

\subsection{\label{sec:level2}Quantum Kicked Top}

The focus of our study are the Floquet eigenvectors $\ket{\lambda_i}$ and eigenphases $\lambda_i$ for the QKT, defined by $\hat{U}\ket{\lambda_i}=e^{-i\lambda_i}\ket{\lambda_i}$, where $\lambda_i$ is defined modulo $2\pi$. Expanding an arbitrary initial state $\ket{\psi(0)}$ in this eigenbasis then gives
\begin{equation}
    \hat{U}^n\ket{\psi(t=0)} =  \ket{\psi(n\tau)}=\sum_i {e^{-i n \lambda_i }}\braket{\lambda_i | \psi(0)}  \ket{\lambda_i},
\end{equation}
We start by noting that the structure of the Hamiltonian is such that it is evident that the dynamics are invariant under $\kappa \mapsto \kappa + 4j\pi$, which immediately restricts our attention to that finite range in general. Of particular interest are kick strengths at which the Floquet eigenvalues are commensurate, lying on the roots of unity up to
a common phase. At such points, there exists a positive integer $q$ satisfying
\begin{equation}
    \hat U(\kappa)^q = e^{i\alpha} I,
\end{equation}
where $\alpha$ is a global phase. We define the order $q$ as the smallest such positive integer. The system then exhibits exact quantum revivals after every $q$ kicks, returning any
initial state to itself up to a global phase. 
Here, we take $j=2$, corresponding to $N=2j=4$ spin $1/2$ particles in the spin permutation symmetric subspace (SPSS) which has dimension $N+1=5$.
Since the Floquet operator was constructed purely from angular momentum operators, it commutes with the total angular momentum operator, $[\hat{U}, \mathbf{\hat{J^2}}] = 0$, ensuring we stay in the SPSS of the the larger $2^N$ dimensional Hilbert space $\mathcal{H}$. 

For the purposes of our study below, we parametrize the initial states by spin-coherent states corresponding to points $(\theta,\phi)$ on the Bloch sphere. 
\begin{equation}
    \ket{\theta,\phi}_N = \left( \cos\left(\frac{\theta}{2}\right)\ket{\uparrow} + e^{i\phi}\sin\left(\frac{\theta}{2}\right) \ket{\downarrow} \right)^{\otimes N}.
\end{equation}
Expanding this yields
\begin{equation}
\begin{split}
\ket{\theta,\phi}_N
&=
\sum_{r=0}^{N}
\sqrt{\binom{N}{r}}
\left(\cos\frac{\theta}{2}\right)^r
\left(\sin\frac{\theta}{2}\right)^{N-r} \\
&\qquad \times e^{i(N-r)\phi} |j,-j+r\rangle,
\end{split}
\label{eq:spin-coherent}
\end{equation}
where $\ket{j,m}$ denotes a Dicke state with collective projection $m$. This gives a natural parametrization of separable initial states over which we build the entanglement landscape. It is also useful to view the kicked top as a periodically driven all-to-all Ising model. Writing the collective spin operator as 
\begin{equation}
    \hat{J_z} = \frac{1}{2} \sum_{r=1}^N \sigma_z^{(r)}.
\end{equation}
gives
\begin{equation}
    \hat{J_z^2} = \frac{N}{4} \hat{I}+ \frac{1}{2}\sum_{r<s}^N \sigma_z^{(r)}\sigma_z^{(s)}.
\end{equation}
Thus, up to an additive constant, the nonlinear $\hat{J_z^2}$ corresponds to an equal strength Ising coupling between all pairs of spins.

We note there that the Floquet operator possesses a symmetry that allows it to be block diagonalized \cite{Dogra_2019}. First, we define the parity indicator $a = N\bmod{2}$ such that $a$ is $1$ for odd $N$ and $0$ otherwise. We then consider the parity operator,
\begin{equation}
    P = i^ae^{-i\pi J_y},
\end{equation}
Its action on the Dicke basis is given by $P\ket{j,m} = i^a(-1)^{j-m} \ket{j,-m}$.
Since $P$ is unitary and that a rotation through $2\pi$ satisfies 
\begin{equation}
    e^{-i2\pi J_y} = (-1)^N I
\end{equation}
it follows that $P^2 = I$ and therefore $P$ is a hermitian operator with eigenvalues $p=\pm 1$.
From its definition, $P$ must commute with $\hat{J_y}$ and commutation with $\hat{J_z^2}$ follows from the fact that
\begin{equation}
    P J_z P^{\dagger} = - J_z
\end{equation}

The Floquet operator must therefore commute with  parity operator $P$. Consequently, the Hilbert space decomposes into the invariant parity sectors $\mathcal{H} = \mathcal{H}_+ \oplus \mathcal{H}_-$
and the Floquet operator can be block diagonalized 
\begin{equation}
    U \cong U_+ \oplus U_-
\end{equation}
One can similarly show that the even-qubit kicked top system has a further reduction of the positive parity sector by introducing operator $Q= e^{-i\pi J_z}$ which has action $Q\ket{m} = (-1)^{m} \ket{m}$ producing a $2+2+1$ block diagonalization for $N=4$, consistent with the decomposition obtained by Dogra \textit{et al.} \cite{Dogra_2019}.

\section{Spectral Degeneracies and entangement landscape sensitivity}

\textbf{Locating Floquet Degeneracies:}
For two polynomials $p(x)$ and $q(x)$, the resultant $\operatorname{Res}(p,q)$ vanishes iff p and q share a common root. It is the determinant of the Sylvester matrix of the two polynomials. For the Floquet operator $U(\kappa)$, we define the characteristic polynomial  
\begin{equation}
    p(\lambda;\kappa)=\det\!\left[\lambda I-\hat{U}(\kappa)\right]= \prod_{i=1}^{N+1}(\lambda - \lambda_i(\kappa)).
\end{equation}
Noting that an eigenphase degeneracy corresponds to a repeated root of $p$, the degeneracies may therefore be located from 
\begin{equation}
    \Delta(\kappa)\equiv\operatorname{Res}(p,\frac{\partial p(\lambda;\kappa)}{\partial \lambda}) = 0
    \label{eq:discriminant}
\end{equation}
where we have used that the resultant of a polynomial and its derivative is proportional to its discriminant. Since $p$ is a monic polynomial of degree five in the present case, the two quantities coincide. Floquet degeneracies therefore occur at values of $\kappa$ for which $\Delta(\kappa)=0$.

\textbf{Entanglement Landscape:}
The bipartite entanglement between one qubit and the remaining $N-1$ qubits is quantified using the linear entropy of the single-qubit
reduced density matrix,
\begin{equation}
    S_Q = 1-\mathrm{Tr}\left(\rho_1^2\right),
\end{equation}
where
\begin{equation}
    \rho_1 = \mathrm{Tr}_{2,\ldots,N}(\rho)
\end{equation}
is obtained by tracing over the remaining $N-1$ qubits. Since the system is restricted to the symmetric subspace, we can also write the entropy as 
\begin{equation}
    S_Q(n) = \frac{1}{2} \left[ 1-\frac{1}{j^2} (\braket{\hat{J}_x}_n^2 +\braket{\hat{J}_y}_n^2 +\braket{\hat{J}_z}_n^2 ) \right].
\end{equation}
We then take average over time to get
\begin{equation}
    \overline{S_Q}(\kappa, \theta,\phi)= \frac{1}{n_{\text{kicks}}} \sum_{i=1}^{n_{\text{kicks}}} S_Q(n;\kappa, \theta,\phi).
\end{equation}
which we can use to define an entanglement landscape
\begin{equation}
    (\theta,\phi) \longmapsto \overline{S_Q}(\kappa, \theta,\phi).
\end{equation}
In what follows, we write $S_Q(\kappa,\theta,\phi)$ for this time averaged quantity, with $n_{\mathrm{kicks}}=200$ or $2000$ as specified in each figure. Using the larger number of kicks yields better-resolution of features but does not qualitatively change any of our conclusions. We use angle brackets $\langle \cdot \rangle$ to denote an addition average over the sphere.
\begin{figure}[tb]
    \centering
    \includegraphics[width=\columnwidth]{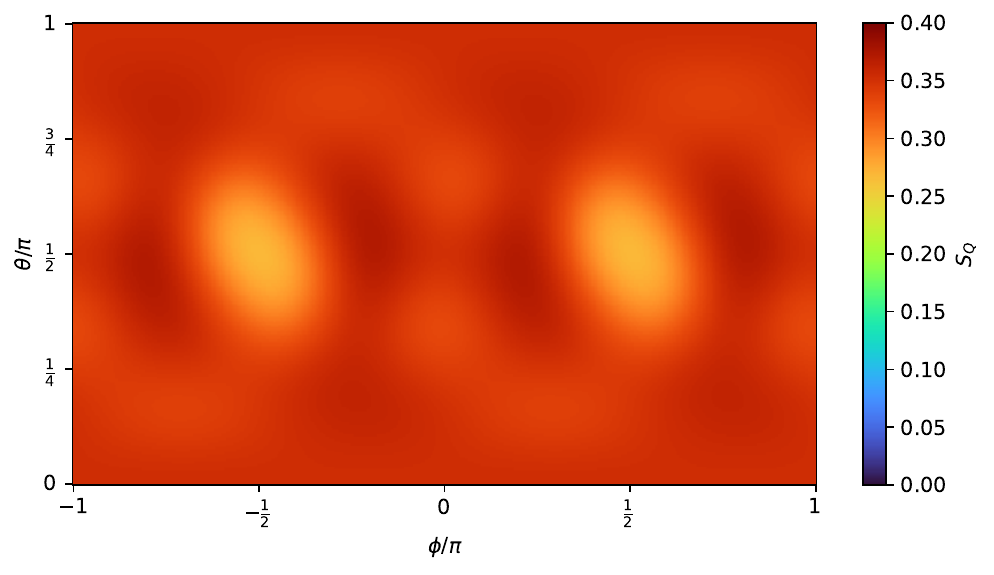}
    \caption{Representative time-averaged entanglement landscape $S_Q(\theta,\phi)$ for 200 kicks for the four qubit kicked top at $\kappa=1.3\pi$. Each point $(\theta,\phi)$ specifies an initial coherent state and its colour denotes the linear entropy averaged over 200 kicks.}
    \label{fig:representative_landscape}
\end{figure}

\textbf{Entanglement Landscape Sensitivity:}
We also define the sensitivity measure $\chi_S$ as

\begin{equation}
    \chi_S(\kappa) = \left[ \frac{1}{4\pi} \int \left(  \frac{\partial S_Q}{\partial (\kappa /\pi)}\right)^2 d\Omega \right]^{\frac{1}{2}}
    \label{eq:chi-s-defn}
\end{equation}

Thus, $\chi_S$ is the root-mean-square change in the time averaged entanglement landscape with respect to kick strength over the sphere of initial coherent states. 
The integration is performed using the solid angle element $d\Omega = \sin\theta\, \mathrm{d}\theta\, \mathrm{d}\phi$ so that equal areas of the sphere are weighted equally, avoiding over measure near the poles. Using this, we define the local contribution to the integral, $\mathcal{I}$ as
\begin{equation}
    \mathcal{I}_S(\theta,\phi;\kappa) = \left( \frac{\partial S_Q}{\partial (\kappa /\pi)} \right) ^2 \sin \theta.
    \label{eq:chi-s-local}
\end{equation}

\section{Results and Discussion}

\begin{figure}[tb] 
    \centering
    \includegraphics[width=\columnwidth]{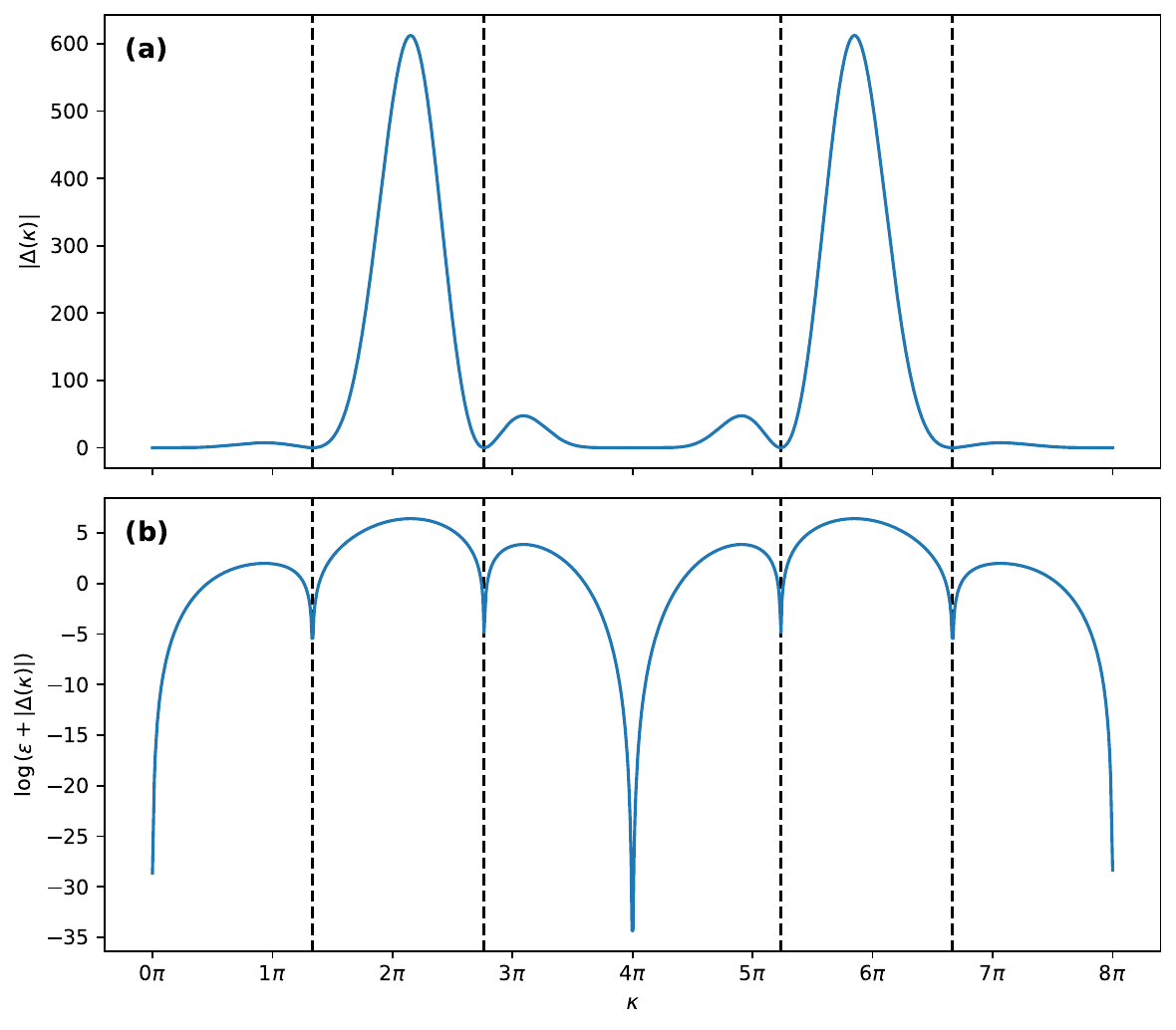}
    \caption{Spectral discriminant of the four-qubit kicked top as a function of $\kappa/\pi$. 
    (a) Magnitude of the discriminant $|\Delta(\kappa)|$
    (b)$\log(\epsilon +|\Delta(\kappa)|)$ with $\epsilon = 10^{-15}$ introduced to regularize the logarithm at $|\Delta(\kappa)|=0 $. 
    Zeros of the discriminant correspond to crossings of Floquet eigenphases and hence to spectral degeneracies. 
    Dashed vertical lines mark the non trivial degeneracies discussed in the text. }
    \label{fig:discriminant}
\end{figure}

\begin{figure}[tb]
    \centering
    \includegraphics[width=\columnwidth]{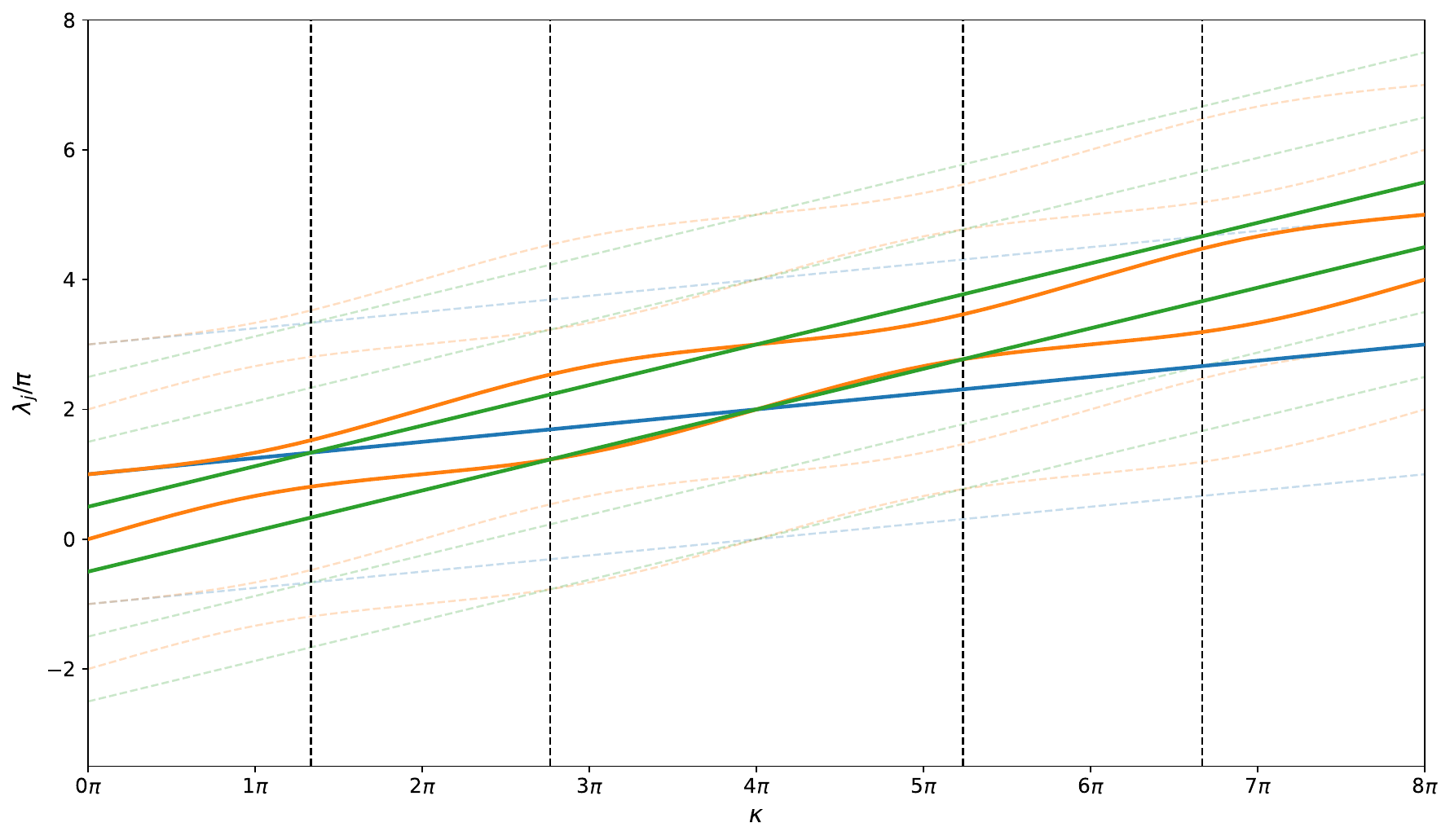}
    \caption{Floquet eigenphases of the four-qubit kicked top as a function of $\kappa/\pi$.
    The eigenphases, defined modulo $2\pi$, are plotted as continuously unwrapped branches with color grouping by invariant subspace. Faint periodic replicas at $\lambda_i \pm 2\pi$ are included to capture all degeneracies.
    Dashed vertical lines mark the non trivial degeneracies discussed in the text. 
    }
    \label{fig:eigenphases}
\end{figure}

\textbf{Spectral structure of the $4$-qubit QKT:}
We first examine the Floquet spectrum and locate its degeneracies using the
discriminant defined in Eq~\ref{eq:discriminant}. Figure~\ref{fig:discriminant} shows
$|\Delta(\kappa)|$ over the full periodic range of the kick strength.
The corresponding Floquet eigenphases are shown in
Fig.~\ref{fig:eigenphases}. The zeros of the discriminant coincide with zero
crossings of the eigenphase branches, confirming that the resultant
correctly identifies the spectral degeneracies. We note that
we did not find any avoided crossings, which can also be found using this technique, when they are present \cite{bhattacharya2007}.

At $\kappa=0$ the kick vanishes, while at $\kappa=4\pi$ it reduces to a rigid $z$-rotation for $j=2$; both cases give degenerate Floquet spectra. In addition to these special-point degeneracies, we identify two non trivial degeneracies at $\kappa=4\pi/3$ and $\kappa=\kappa^*$ where we define $\kappa^*$ to be
\begin{equation}
\begin{aligned}
\kappa^* &=
8\arcsin\left[
\sqrt[3]{\frac{9+\sqrt{57}}{72}}
+
\sqrt[3]{\frac{9-\sqrt{57}}{72}}
\right] \\
&\approx 2.7648\pi
\end{aligned}
\end{equation}
The other non trivial degeneracies can be found via the following symmetry argument. 
In the $\hat{J_z}$ basis, the matrix representation of $\hat{J_z^2}$ is real, while $\hat{J_y}$ is purely imaginary, so 
\begin{equation}
    \hat{U}(-\kappa) = \hat{U}(\kappa)^*
\end{equation}
Hence its characteristic polynomial satisfies 
\begin{equation}
\begin{aligned}
p(\lambda,-\kappa)
&=
\det\!\left[\lambda I-\hat{U}(-\kappa)\right] \\
&=
\det\!\left[\lambda I-\hat{U}(\kappa)^*\right] \\
&=
\det\!\left[\lambda^* I-\hat{U}(\kappa)\right]^* \\
&=
p(\lambda^*,\kappa)^* .
\end{aligned}
\end{equation} 
and therefore 
\begin{equation}
    \operatorname{spec}\hat{U}(-\kappa) = \left[  \operatorname{spec}\hat{U}(\kappa)    \right]^*.
\end{equation}
combining this with the $8\pi$ periodicity of $\hat{U}(\kappa)$, degeneracies at $\kappa^*$ are accompanied by degeneracies at $8\pi - \kappa^*$, giving reflection symmetry about $\kappa = 4\pi$.

Local maxima of $|\Delta(\kappa)|$ were also examined, but showed no systematic correspondence with either spectral degeneracies or enhanced entanglement landscape sensitivity. We therefore do not consider them further.

\textbf{Mapping spectral structure to entanglement dynamics:} 
Ref.~\citenum{Bhosale2018} demonstrated that although the Floquet operator itself is $4j\pi$-periodic, correlation dynamics are $2j\pi$-periodic. The
reason for this shorter period comes from the fact that when $\kappa=2j\pi$, the kick operator becomes
\begin{equation}
    U_{\mathrm{kick}} = \exp\left(-i\pi J_z\right)
\end{equation}
and on the SPSS, they show that $\exp\left(-i\pi J_z\right)=\otimes\sigma_z$, and therefore acts as local unitaries. Entanglement dynamics are insensitive
to action by local unitaries, and therefore the entanglement dynamics must be $2j\pi$-periodic. When combined with the fact that the Floquet operator
is reflective-symmetric about $\kappa=4\pi$, Ref.~\citenum{Bhosale2018} shows that the entanglement dynamics exhbit reflection symmetry about $j\pi$, 
meaning that unique entanglement dynamics only occur on the interval $[0,2\pi]$ for $j=2$. This means that spectral features on the interval $[2\pi,4\pi]$ are
folded back onto the interval $[0,2\pi]$, and thus we see a spike in the entanglement sensitivity at $\kappa=4\pi-\kappa^*$, even though there is no 
eigenpphase degeneracy at that point.  Likewise, there is a spike in the entanglement sensitivity at $\kappa=4\pi-4\pi/3$, again even though there is no eigenphase degeneracy at that point either. 

\textbf{Finite-Order Floquet Dynamics and Quantum Revival:}

For the $N=4$ kicked top, the Floquet operator exhibits finite order dynamics at particularly low order. These cases occur at $\kappa=0,2\pi,4\pi$, with corresponding orders $q=4,8,2$, respectively. In each case, $\alpha=0$, resulting in exact quantum revivals after $q$ kicks.  Two of 
these cases, $\kappa=0,4\pi$ do not produce entanglement for any number of kicks,
so $ S_Q(0)= S_Q(4\pi)=0$ for any initial coherent state and so their sphere averages vanish identically.  For $\kappa=2\pi$, the
entanglement dynamics have a period of 4 kicks, and so for every fourth kick,
the entanglement goes to 0.  
Because of this, $\langle S_Q(2\pi)\rangle\approx 3/4 \langle \langle S_Q \rangle \rangle _{\kappa}
$ where $ \langle \langle S_Q \rangle \rangle _{\kappa} $ is the entanglement averaged also over all $\kappa$ values.

The values $\kappa=\pi,3\pi$ also exhibit finite order dynamics, corresponding to the
48th roots of unity, and return to zero entanglement every 24 kicks.  
Indeed $\langle S_Q(\pi)\rangle$ exhibits a clear dip when we average over 
the first 24 kicks.  
However, as the averaging time increases, increasingly fine oscillatory structure develops in the neighborhood of $\kappa = \pi$ (see 
fig.~\ref{fig:S_Q_zoom_in_on_pi} in appendix~\ref{app:longer-periodic-orbits}).
This behavior near $\kappa=\pi$ may be
understood in terms of the accumulation of small deviations
from the exact periodic dynamics. At $\kappa=\pi$, the
Floquet operator satisfies $U(\pi)^{48}=I$. For nearby
values of $\kappa$, we therefore have
$U(\pi+\delta\kappa)^{48}=I+O(\delta\kappa).$
Although these deviations are initially small, they can
accumulate over successive periods, leading to an
increasingly sensitive dependence of the entanglement on
$\kappa$. By 2000 kicks $\langle S_Q(\kappa)\rangle$ appears comparitively
smooth in the neighborhood of $\kappa=\pi$
suggesting the expected suppression by a factor of 
23/24 is not significant in comparison with other effects.

\textbf{Parametric sensitivity of the entanglement landscape:}
Figure~4 shows the average entanglement $\langle S_Q\rangle$ (top) and entanglement landscape susceptibility $\chi_S$ (bottom) defined in Eq.~15 over the range $0\leq\kappa\leq4\pi$. For $j=2$, the underlying entanglement measure is $4\pi$-periodic and reflection symmetric about $\kappa=2\pi$ \cite{Bhosale2018}, symmetries which are inherited by $\chi_S$.
Owing to the symmetry about $2\pi$, there are four unique spikes in the entanglement
sensitivity. These four spikes fall into three overlapping categories. The spikes at
$\kappa=0,4\pi/3,\kappa^*$ correspond to Floquet eigenphase degeneracies.  The spikes
at $\kappa=0,2\pi$ correspond to the Floquet operator being periodic. Finally
the spikes at $\kappa=4\pi/3,2\pi$ are examples of nonlinear resonances, since they
are given by $(r/s)4\pi j$. They are fairly low-order resonances, with $(r/s)=(1/4),(1/6)$.
If we include the mirror-symmetric points of $4\pi$ and $8\pi/3$ we also obtain 
$(r/s)=(1/2),(1/3)$. Intriguingly $(1/5)$ is the only low-order resonance absent from this
pattern.

The sensitivity to changing $\kappa$ can come from several different sources. Three that we have identified as being useful in understanding the metric are the change to the average entanglement, $S_Q$, change to the standard deviation $\sigma_{S_Q}$, or
changes to the locations of bumps on the surface. See Appendix~\ref{app:entanglement-analysis} for
a more detailed break-down of these different effects.
\begin{figure}[tb]
    \centering
    \includegraphics[width=\columnwidth]{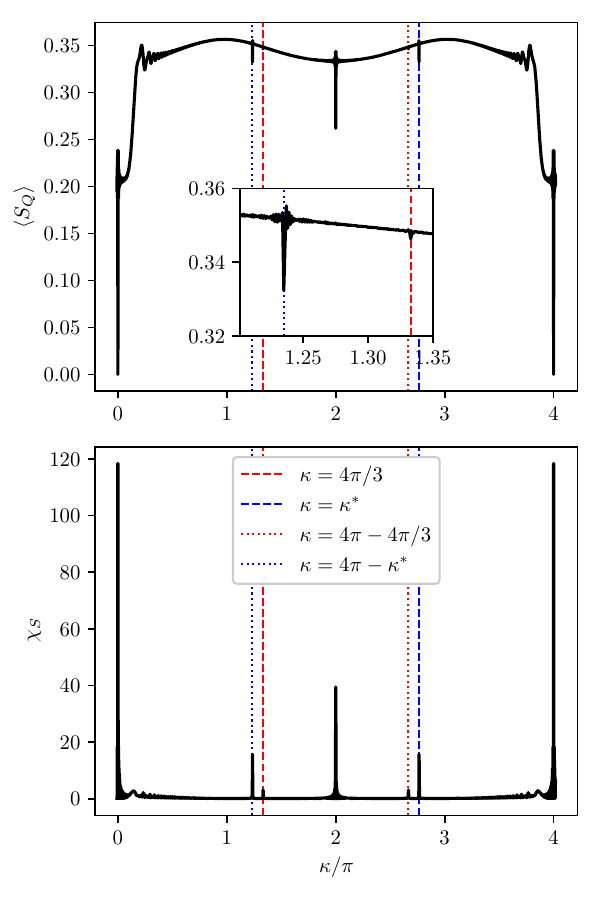}
    \caption{Average entanglement $\langle S_Q\rangle$ (top) and entanglement landscape susceptibility $\chi_S$ (bottom) for 2000 kicks as a function of $\kappa/\pi$ over one full $4\pi$ period. The derivative was approximated using a centred finite difference  with steps of $\Delta( \kappa/ \pi ) = 0.001$ . Vertical lines indicate the non trivial degeneracies and their reflection as described in the text. The inset enlarges the region 
    $1.2\leq\kappa/\pi\leq1.4$ around the degeneracy point $\kappa = 4/3 \pi$ and the reflection point $\kappa = 4 \pi -\kappa^*$. Each spike in the entanglement sensitivity corresponds to a dip in the overall average entanglement. However, these dips cannot fully explain the spikes at the 
    nontrivial degeneracy points, where shape change is the more important feature (see Appendix~\ref{app:entanglement-analysis} for further analysis).}
    \label{fig:chiS}
\end{figure}

\textbf{Spectral origins of entanglement landscape sensitivity:}
The connection between Floquet spectral structure and entanglement landscape sensitivity follows by expanding the spin expectation values in the Floquet eigenbasis:
\begin{equation}
\langle J_\nu\rangle_n
=\sum_{a,b}c_a^*c_b J^\nu_{ab}
e^{in(\lambda_a-\lambda_b)},
\end{equation}
where $c_a=\langle\lambda_a|\theta, \phi\rangle$,
$J^\nu_{ab}=\langle\lambda_a|J_\nu|\lambda_b\rangle$.
The linear entropy averaged over $T$ kicks is then
\begin{equation}
S_Q^{(T)}(\Omega;\kappa)
=\frac12-\frac{1}{2j^2}
\sum_{a,b,c,d}C_{abcd}(\Omega,\kappa)
K_T(\omega_{abcd}),
\end{equation}
with
\begin{equation}
\begin{aligned}
C_{abcd}
&=c_a^*c_b c_c^*c_d
\sum_{\nu=x,y,z}J^\nu_{ab}J^\nu_{cd},\\
\omega_{abcd}
&=\lambda_a-\lambda_b+\lambda_c-\lambda_d,\\
K_T(\omega)
&=\frac1T\sum_{n=1}^{T}e^{in\omega}
=e^{i(T+1)\omega/2}
\frac{\sin(T\omega/2)}{T\sin(\omega/2)}.
\end{aligned}
\end{equation}
All initial-state dependence resides in $C_{abcd}$, while
$K_T$ selects the interference terms that survive time averaging.
At resonance, $\omega=0$ modulo $2\pi$, $K_T=1$; otherwise,
\begin{equation}
|K_T(\omega)|\leq
\min\left\{1,\frac{1}{T|\sin(\omega/2)|}\right\}.
\end{equation}
We note that time averaging suppresses off-resonant terms, while $|K_T|$ approaches unity within a frequency window of order $1/T$ around $\omega=0$ modulo $2\pi$. For the averaging time used here ($T=200$), this window is narrow.

\begin{figure}[tb]
    \centering
    \includegraphics[width=\columnwidth]{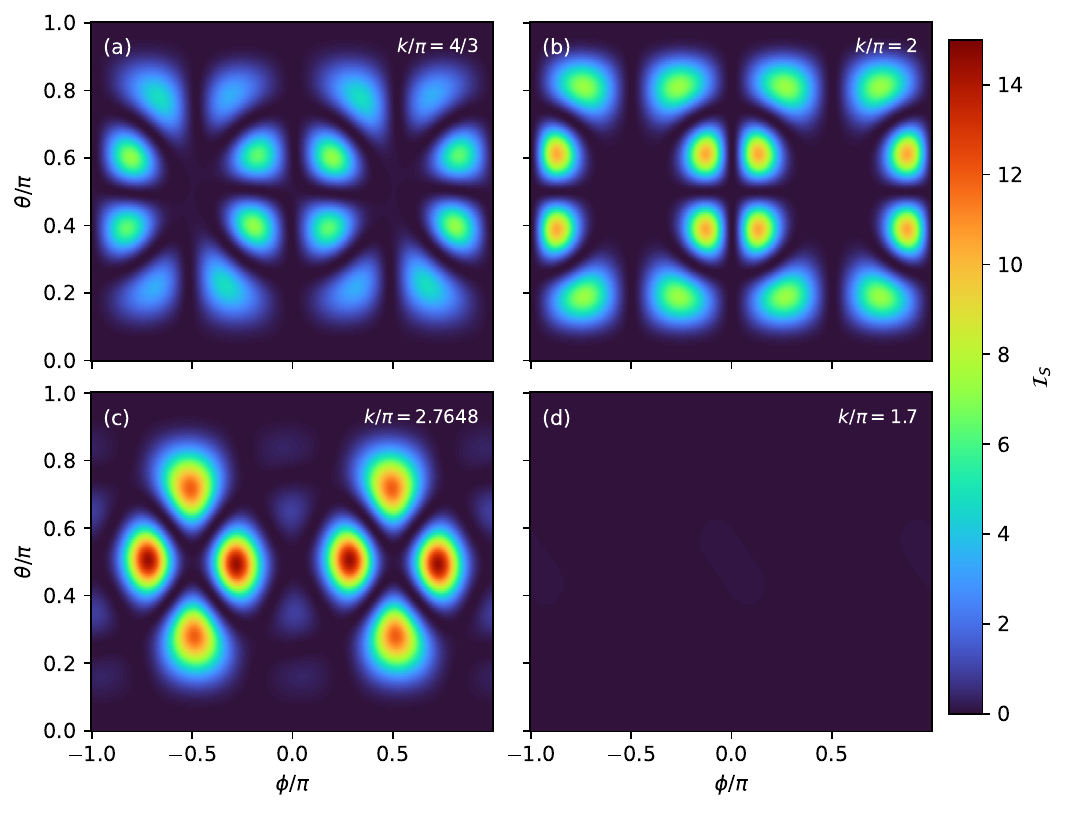}
    \caption{Initial-state dependence of the local contribution to the entanglement landscape susceptibility $\chi_S$ for 200 kicks at two Floquet degeneracies  
    (a) $\kappa/\pi=4/3$ and (c) $\kappa/\pi=2.7648$ compared with (b) the notable nondegenerate point $\kappa/\pi=2$ and (d) a generic degenerate point $\kappa/\pi=1.7$.
    Brighter regions indicate coherent states for which entanglement is most sensitive to changes in $\kappa$}
    \label{fig:chiS_contribution}
\end{figure}

We point out that degeneracies and evenly spaced eigenphases can admit more combinations satisfying $\lambda_a+\lambda_c=\lambda_b+\lambda_d$ modulo $2\pi$ than a generic spectrum. These additional resonances allow further contributions, weighted by the angular-dependent coefficients $C_{abcd}(\Omega,\kappa)$, to survive time averaging. As $\kappa$ varies, these terms can enter or leave the resonance window, producing rapid landscape deformation, subject to their weights and cancellations. This mechanism is consistent with the rapid deformations observed near $\kappa=4\pi/3$, $2\pi$, and $\kappa^\ast$ in the supplementary animation, and quantified in
fig.~\ref{fig:S_Q_changing_metrics} in Appendix~\ref{app:entanglement-analysis}. This interpretation is also consistent with Fig.~\ref{fig:chiS_contribution}, where the spectrally special kick strengths exhibit pronounced angular structure in the local contribution to $\chi_S$, compared with the weak response at the generic point.

\begin{figure}[tb]
    \centering
    \includegraphics[width=\columnwidth]{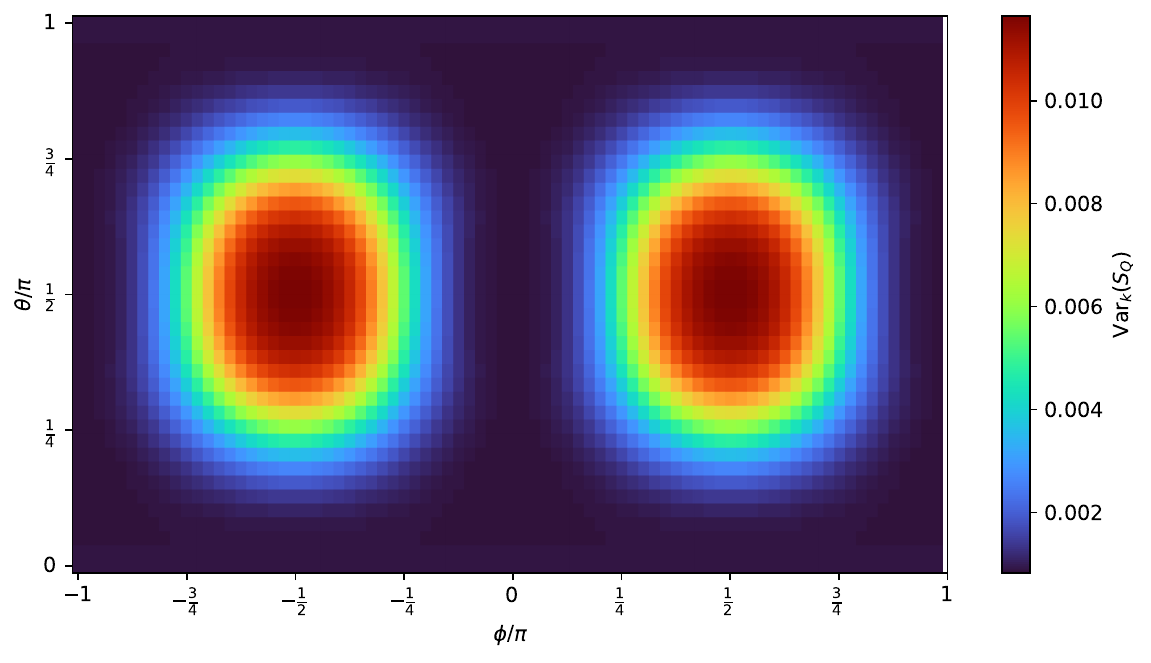}
    \caption{Finite-time variance of the time-averaged entanglement measure for 200 kicks over the sphere of coherent initial conditions. Brighter color corresponds to regions on the sphere where the spread of entanglement values is largest. }
    \label{fig:variance}
\end{figure}

\textbf{Initial condition dependence:}
Here we examine the initial state dependence of the local contribution to $\chi_S$, defined in Eq.~\ref{eq:chi-s-local} and seen in Fig.~\ref{fig:chiS_contribution}. We see that the plot shows highly symmetric regions of large change localised on specific initial conditions. We see that for kick strengths around degeneracies, the local contribution $\mathcal{I}_S$ has large peaks, whereas for a typical initial condition, the fluctuations in entanglement landscape around this point are minimal. 
We also consider the variance in $S_Q$ across points on the Bloch sphere seen in Fig.~\ref{fig:variance}. This is dominated by the points $(\pi /2,\pi /2)$ and $(-\pi /2,\pi /2)$ due to their large decay around $\kappa = 0 $.

\section{Conclusion}

In conclusion, we have shown that for a time-periodic many-body quantum system, the spectral structure as visible in the spectral discriminant of the Floquet operator shows cusps which we identify as corresponding to parameter locations where the quantum many-body dynamics (where we have focused here on finite-time entanglement dynamics analyzed with various different metrics) are either unusual in being periodic or are particularly sensitive to variation in parameter. We show that these unusual parametric locations include but go beyond those corresponding to `primary quantum resonances' where the spectra of the two competing terms in the Hamiltonian can be understood to be in generic `resonance' with each other. We note that although we have not explicitly simulated the impact of environmental decoherence, etc, the finite-time perspective shows that all of the reported results are well within the capability of current experiments. 

Further, while the bulk of our study occurs in $j=2$, i.e. the $4$-qubit system, since the resonance phenomena as well as periodicity in $\kappa$ must strictly apply independent of $j$ we also consider how the behavior changes with increasing $j$ and in particular the  limit of large $j$. We find that the number-density in parameter space of these spectral cusps and dynamically unusual parameter values increases with the number of qubits, indicating far more complicated parametric dependence in non-linear quantum systems than can be deduced by looking at the classical limit.  A similar perspective using quantum eigenfunctions should allow for the unfolding of uniquely quantum initial-condition dependence independent of the behavior of the nonlinear classical limit. Thus, searching for spectral features -- such as avoided crossings and even spacings -- beyond the usual `spectral statistics' in this non-linear Hamiltonian systems has helped us identify unusual dynamical regimes and parameter dependence (`a bifurcation diagram') beyond the usual semiclassical perspective of quantum mechanics displaying order-to-chaos transition. Beyond this broader perspective, these techniques should also be of value in engineering quantum systems with desirable properties related to parametric sensitivity such as for sensors, for example. 

\begin{acknowledgments}
AKP would like to thank Amin Nizami for a very helpful discussion about quantum resonances. He would also like to acknowledge Noah Pinkney, Alex Kiral, Sudheesh Srivastava, Ryan Quinn, and Alex Gran for previous work on understanding these issues using a different version of these tools and metrics that did not prove as fruitful.
\end{acknowledgments}

\begin{appendix}

\section{Entanglement analysis\label{app:entanglement-analysis}}
Here we present a more in-depth analysis of the entanglement landscape as a function of
$\kappa$.  For each value of $\kappa$, we use pyshtools to decompose $S(\theta,\phi)$ 
into spherical harmonics \cite{pyshtools}:
\begin{equation}
    S(\theta,\phi)=\sum_m a_{l,m} Y_{l,m}(\theta,\phi),
\end{equation}
and then compute the power spectrum $S_l=\sum_m c_{lm}^2$, where the mean is
$\langle S_Q\rangle=S_0$ and the variance is $\sigma_{S_Q}^2=\sum_{l>0}S_l$. We also 
compute $l_\mathrm{eff}=\sum_l l(l+1)S_l/\sigma^2_{S_Q}$, which quantifies how 
bumpy the surface is. To find the minimum and maximum values we evaluate the function
on a grid.
\begin{figure}[htbp]
    \centering
    \includegraphics[width=\columnwidth]{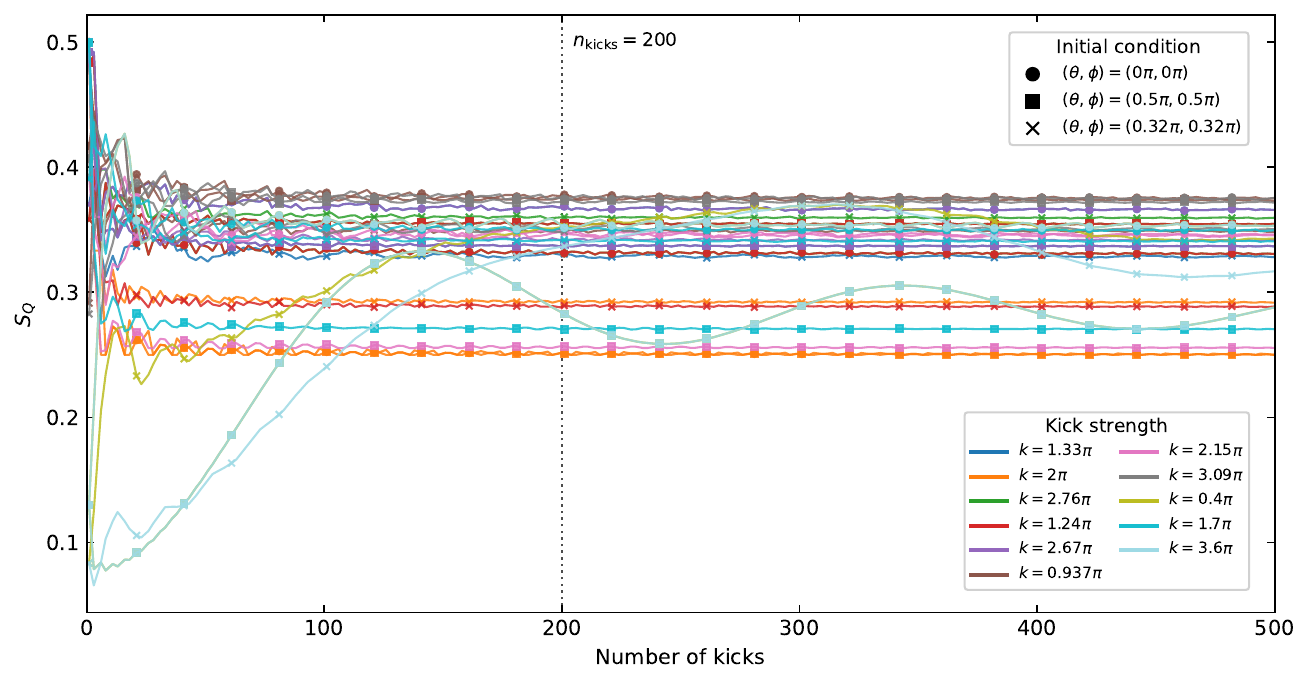}
    \caption{Convergence of the time-averaged entanglement measure with the number of kicks.}
    \label{fig:SQ_convergence}
\end{figure}

\begin{figure}[htbp]
    \centering
    \includegraphics[width=\columnwidth]{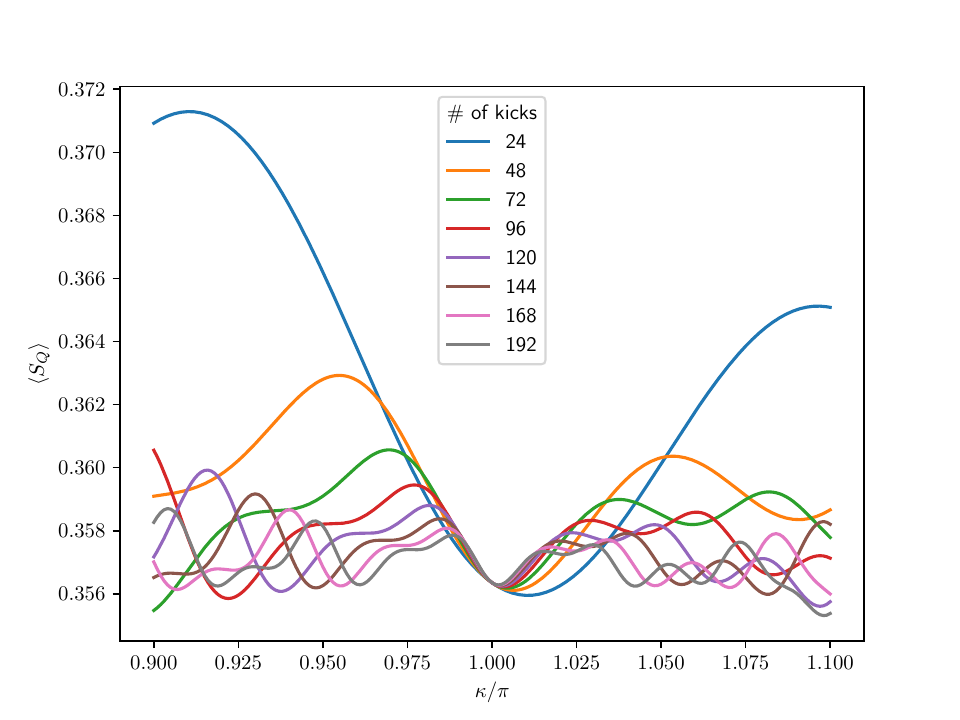}
    \caption{Shows the time- and space-averaged entanglement for $\kappa$ near $\pi$,
    for small multiples of 24 kicks. For exactly 24 kicks, there is a clear dip near 
    $\pi$.  However, as the number of kicks increases, other effects bring the total
    entanglement down at neighbouring points.  By 2000 kicks, the average entanglement
    is nearly flat.}
    \label{fig:S_Q_zoom_in_on_pi}
\end{figure}

\begin{figure}[tb]
    \centering
    \includegraphics[width=\columnwidth]{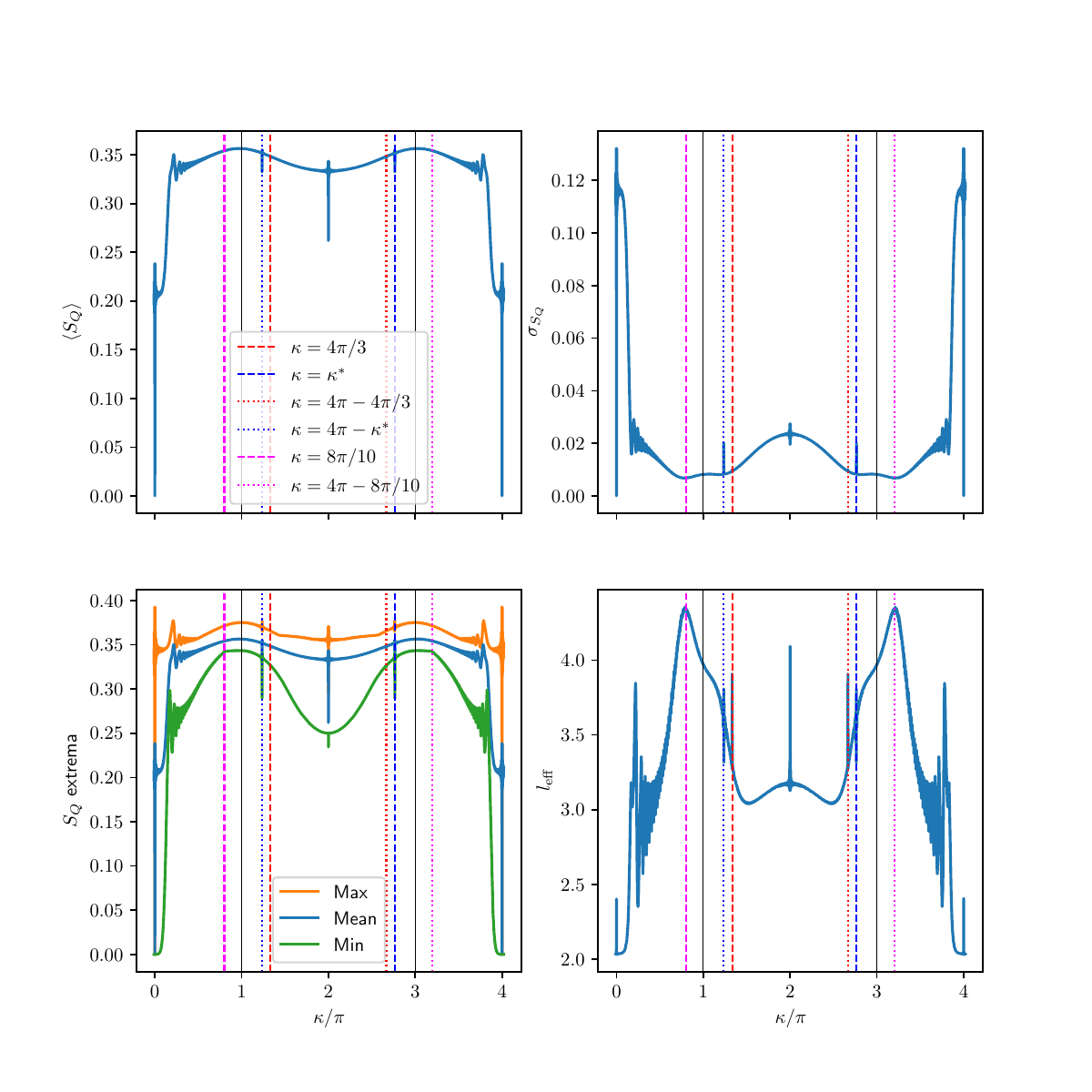}
    \caption{Characterization of the entanglement landscape as a function of $\kappa$, 
    averaged over 2000 kicks. Upper-left panel shows entanglement landscape averaged 
    over all initial conditions. Upper-right panel shows the standard deviation of the 
    entanglement. Lower-left panel shows the mean again, along with the minimum and maximum
    entanglement. Lower-left panel shows $l_\mathrm{eff}$, which is the effective spherical 
    harmonic order needed to approximate the surface roughness.}
    \label{fig:S_Q_analysis}
\end{figure}

As was shown in the main text, and reproduced here in Fig.~\ref{fig:S_Q_analysis},
the average entanglement $\langle S_Q\rangle$ dips at each value of $\kappa$ where $\chi_S$
peaks.  These dips are particularly pronounced at $0,2\pi,4\pi$, because these values of 
$\kappa$ correspond to locations where the entanglement is either always 0 ($0,4\pi$), or
returns to zero every fourth kick ($2\pi$).  There are also dips at the other two spikes,
$\kappa=4\pi/3,\kappa^*$.  These two points correspond to zero-crossings of the eigenphase
spectrum.  The dips in the mean are primarily caused by dips in the minimum value of 
entanglement, rather than dips in the maximum values of entanglement, and there is therefore
a corresponding small spike in the standard deviation of $S_Q(\theta,\phi)$.  Zero crossings
result in the affected subspace of the Hamiltonian becoming stationary. A stationary subspace
will protect entanglement that already exists, but also prevent its formation. Since our initial
conditions are unentangled, this seems to result in decreased entanglement, particularly at
certain initial conditions.

As we noted in the main text, although $\kappa=\pi,3\pi$ are periodic orbits (period 48, specifically), 
and exhibit precisely 0 entanglement every 24 kicks, they show no dip in the entanglement averaged
over $\theta,\phi$ and over sufficiently many kicks (after about 100 or so, the dip disappears). Rather
they lie very close to the maximum average entanglement. We show here that $\kappa=\pi,3\pi$  are local
maxima in the maximum entanglement over $\theta,\phi$, $\max_{\theta,\phi} S_Q(\theta,\phi)$.  The minimum does
not peak at $\kappa=\pi,3\pi$, and neither does the mean.

Another point of note is the location $\kappa=8\pi/10$, which exhibits the global maximum in 
$l_\mathrm{eff}$, which is the effective spherical harmonic order needed to approximate the 
surface roughness.  Thus in some sense this point has the maximum variation in initial condition
$\theta,\phi$, and yet on the other hand it is also very close to the minimum standard deviation.
So although the entanglement surface is more highly structured, it also has the narrowest spread,
which is visibly clear when looking at the minimum and maximum values of entanglement.
Both of the zero crossings are sharp local maxima in $l_\mathrm{eff}$.

We now turn to understanding $\chi_S$ in more detail. There are many contributions to the change in 
the entanglement landscape. We have found that the three most useful ways of understanding the change
are the change in the mean $d\langle S_Q\rangle/d\kappa$, the change in the standard deviation, 
$d\sigma_{S_Q}/d\kappa$, and the change in the ``shape", which we compute by first converting each 
value of $S_Q(\kappa)$ into a unit vector by computing
$u=\left(S_Q-\langle S_Q\rangle\right)/\sigma_{S_Q}$.  
We then compute $|du/d\kappa|$ by finding the angle between each $u$.

\begin{figure}[tb]
    \centering
    \includegraphics[width=\columnwidth]{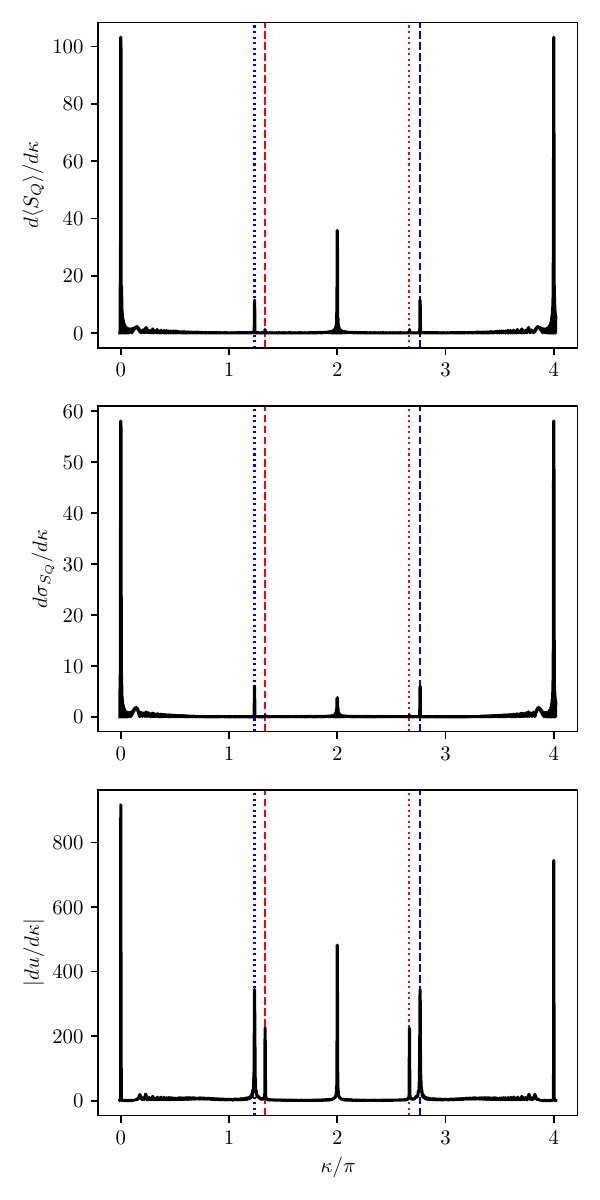}
    \caption{Some of the contributions to the entanglement sensitivity for 2000 kicks. 
    Top panel shows the derivative of the mean with respect to $\kappa$. 
    Middle panel shows the derivative of the standard deviation with 
    respect to $\kappa$. Bottom panel shows the rate of change of the 
    bump locations.}
    \label{fig:S_Q_changing_metrics}
\end{figure}

\section{Longer periodic orbits\label{app:longer-periodic-orbits}}
The shortest periodic orbits of 2 and 4 produce no entanglement in the
SPSS. The period 8 orbit of $2\pi$ does produce entanglement, and the
dynamics return the state to 0 entangelement every 4 kicks. This brings
the average entanglement down to a little more than 3/4 the value obtained
by averaging over all $\kappa$.  As mentioned in the main text, we do not
see an analagous dip at $\kappa=\pi$, even though it returns to 0 entanglement
every 24 kicks. It seems that a reduction in the entanglement here of 23/24,
due to the periodic return to 0 entanglement, is not a large enough effect to
bring it down relative to neighbouring $\kappa$ points.  We show that here in Fig.~\ref{fig:S_Q_zoom_in_on_pi}.

\section{Numerical convergence\label{app:Numerical convergence}}
Finally, we report on the dependence on the averaging time $n_{\mathrm{kicks}}$, as shown in Fig.~\ref{fig:SQ_convergence}. While some initial conditions $(\kappa, \theta, \phi)$ appear to converge within 200 kicks, others retain noticeable finite-time variation. Our sensitivity and landscape results should therefore be interpreted as finite-time numerical quantities rather than infinite-time limits. This choice is nevertheless physically motivated, since current experimental realizations of the quantum kicked top probe dynamics over only tens of kicks \cite{Chaudhury2009}.

\end{appendix}

\bibliographystyle{apsrev4-2}
\bibliography{references}

\end{document}